\documentclass{eas}
\usepackage{graphicx}
\DeclareGraphicsExtensions{.eps} 
\usepackage{hyperref}
\usepackage{textcomp}

\begin{document}

\title{Dedicated front-end and readout electronics developments for real time 3D directional detection of dark matter with MIMAC.} 
\runningtitle{O.~Bourrion: MIMAC electronics}
\author{O.~Bourrion}\address{Laboratoire de Physique Subatomique et de Cosmologie,\\ 
Universit\'e Joseph Fourier Grenoble 1,\\
  CNRS/IN2P3, Institut Polytechnique de Grenoble,\\
  53, rue des Martyrs, Grenoble, France}
\author{G.~Bosson}\sameaddress{1}
\author{C.~Grignon}\sameaddress{1}
\author{J.P.~Richer}\sameaddress{1}
\author{O.~Guillaudin}\sameaddress{1}
\author{F.~Mayet}\sameaddress{1}
\author{J.~Billard}\sameaddress{1}
\author{D.~Santos}\sameaddress{1}
\begin{abstract}

A complete dedicated electronics, from front-end to back-end, was developed to instrument a MIMAC prototype. A front end ASIC able to monitor 64 strips of pixels and to provide their individual ``Time Over Threshold'' information has been designed. An associated acquisition electronics and a real time track reconstruction software have been developed to monitor a 512 channel prototype. This auto-triggered electronic uses embedded processing to reduce the data transfer to its  useful part only, i.e. decoded coordinates of hit tracks and corresponding energy measurements. The electronic designs, acquisition software and the results obtained are presented.

\end{abstract}
\maketitle
\section{Introduction}
Directional detection of dark matter is known to be a promising search strategy of galactic Dark Matter (Spergel~\etal~\cite{Spergel}, Ahlen~\etal~\cite{Ahlen}). Recent studies have shown that, within the framework of dedicated statistical data analysis, a low exposure directional detector could lead either to a high significance discovery of galactic Dark Matter (Billard~\etal~\cite{billard.2010a}, Billard~\etal~\cite{billard.2011}) or to a conclusive exclusion (Billard~\etal~\cite{billard.2010b}).



A gaseous micro-TPC matrix, filled with either $\rm ^3He$, $\rm CF_4$ or $\rm C_4H_{10}$ has been developed within the MIMAC project (Santos \etal~\cite{MIMAC}).
To demonstrate the relevance of the concept, specific front-end ASIC and a dedicated acquisition electronic 
were developed in order to equip a prototype detector featuring an anode of $\rm 10.85 \times 10.85\ cm^2$ where 
$2 \times 256$ strips are monitored. This auto-triggered acquisition electronic uses embedded processing to reduce data transfer to its useful part only, 
i.e. decoded coordinates of hit tracks and corresponding energy measurements. 
To be fully exploited, an acquisition software with on-line monitoring and track 
reconstruction has been written.

This paper is organized as follows: section~\ref{detPrinciple} presents the detector and its readout principle, section~\ref{ASIC.sec} describes the front end ASIC developed for MIMAC. Section~\ref{DAQ.sec} describes the acquisition board and the algorithm embedded in the FPGA. Finally, the acquisition software and the on-line 3D reconstruction strategy are presented in section~\ref{soft.sec}. Eventually, a short summary is given in section~\ref{SummarySec}.

\section{MIMAC detector readout principle}
\label{detPrinciple}
\begin{figure}[ht]
\begin{center}
\includegraphics[width=0.65\textwidth]{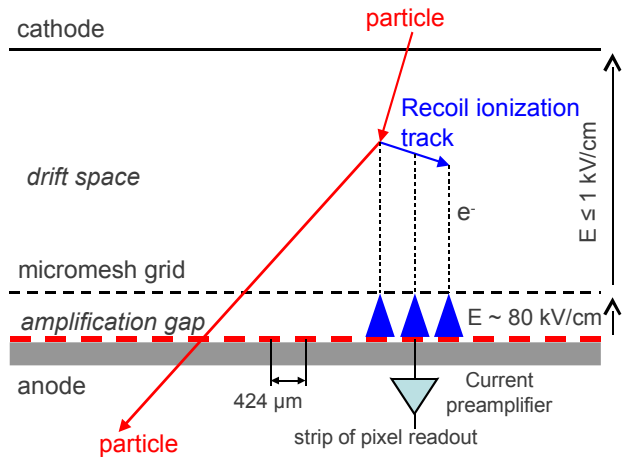}
\caption{Schematic of the MIMAC micro-TPC using a micromegas composed of a pixelized anode featuring 2 orthogonal series of 256 strips of pixels  and a micromesh grid defining the delimitation between the amplification (grid to anode) and the drift space (cathode to grid).}
\label{microTPC}
\end{center}
\end{figure}

As shown in fig.~\ref{microTPC}, the MIMAC prototype \textmu TPC is composed of a pixelized anode featuring 2 orthogonal series of 256 strips of pixels (X and Y) (Iguaz~\etal~\cite{Iguaz}) and a micromesh grid defining the delimitation between the amplification (grid to anode) and the drift space (cathode to grid).
Each strip of pixels is monitored by a current preamplifier and the fired pixel coordinate is obtained by using the coincidence between the X and Y strips (the pixel pitch is 424\,\textmu m). A coincidence is defined as having at least one strip of pixels fired in each direction (X, Y) at the same sampling time.  The ionization energy of the recoil energy is obtained by instrumenting the micromesh grid with a Charge Sensitive Preamplifier (CSP).

\begin{figure}[ht]
\begin{center}
\includegraphics[width=0.85\textwidth]{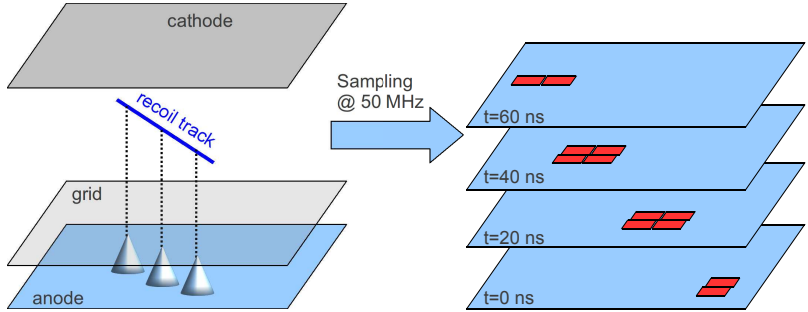}
\caption{The coordinates in the anode plane (X, Y) are reconstructed by collecting primary electrons produced in the drift region. Knowing the electron drift velocity, the third dimension (Z) is obtained by sampling  the anode signal every 20\,ns. Due to the multiplexed readout of the anode, each time slice picture is rectangular.}
\label{ThirdDim}
\end{center}
\end{figure}

As illustrated in fig.~\ref{ThirdDim}, the coordinates in the anode plane (X, Y) are reconstructed by collecting primary electrons produced in the drift region. Knowing the electron drift velocity, the third dimension (Z) is obtained by sampling  the anode signal every 20\,ns. Note, that due to the multiplexed readout of the anode, each time slice picture is rectangular.

\section{Front end ASIC}
\label{ASIC.sec}
\subsection{Requirements}
In the early stage of the project it was decided to design an ASIC in order to be able to fulfill the final objective, which is to equip about 2500 chambers of 1024 strips of pixels (512+512). This minimizes space requirement and power demand and allows cost reduction on a large scale. After going through a first prototype phase, 16 channels ASICs equipping a $2 \times 96$ strips of pixels chamber (Richer~\etal~\cite{Richer}), a 64 channel version was designed. This was determined to be a good balance between integration scale on one side and complexity, fabrication yield and available packages on the other side. 

To be able to recover the third coordinate (Z) of the track, a fast switching current comparator having a threshold as low 200\,nA must be designed in order to have a precise time over threshold measurement of each current preamplifier output. This requirement is driven by the worst case where the recoil energy is as low as 500\,eV in a chamber having its gain  limited to 3000 and where the diffusion is maximized (i.e interaction at the farthest of the anode) and the recoil track is parallel to the anode (the charge deposit is distributed along different strips). Another strong system requirement is to minimize the board level interconnection to allow an easy integration with a readout system.

\subsection{Design details}
\begin{figure}[ht]
 \begin{center}
 \includegraphics[width=0.8\textwidth]{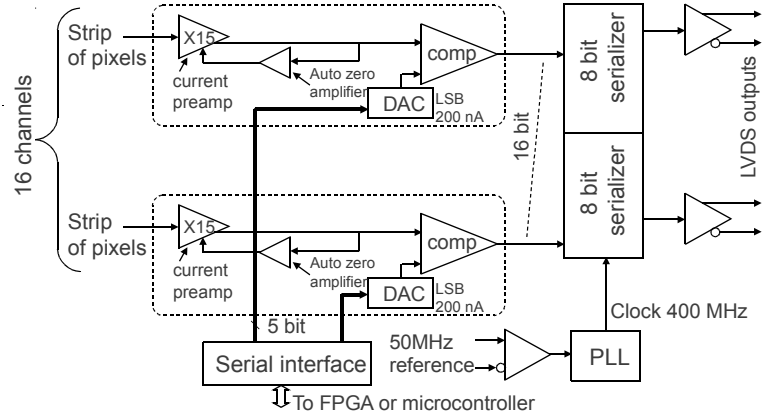}
\caption{Front end ASIC block diagram. The 64 channels are decomposed in 4 groups of 16 channels. Each channel is composed of a current preamplifier having a gain of 15, a fast comparator (modified CMOS inverter kept in linear region) and a 5 bit DAC for setting the threshold. A serial link is used to configure the DAC. The comparator outputs are transferred serially.}
\label{BlockASIC}
\end{center}
\end{figure}

The front end ASIC, whose block diagram is shown in fig.~\ref{BlockASIC}, is composed of 4 groups of 16 channels. Each channel is composed of a current preamplifier having a gain of 15, a fast comparator (modified CMOS inverter kept in linear region) and a 5 bit DAC for setting the threshold. 
A trade-off was made between the DAC resolution and the achievable preamplifier offset. 
The DAC dynamic range, and thus its design complexity can be greatly reduced by using an autozero preamplifier. This kind of amplifier measures periodically (every second) its offset during a few dozen \textmu s and determines the compensation to apply to reduce the residual output offset. 
The 5 bit DAC is designed with a 200\,nA LSB (input equivalent: 13.3\,nA).

The comparator outputs are sampled at a 50\,MHz rate and serialized at 400\,MHz, thereby reducing the interconnection by a factor of 8 and diminishing the power consumption.  The serial outputs rely on the Low Voltage Differential Signaling (LVDS) standard to lower the electronic noise. It should be noted that using the same reference clock allows synchronous sampling between ASICs.

A slow serial link is used to configure the 64 DACs and to individually enable/disable each channel (kill eventual dead channels, ...).

The ASIC was fabricated in AMS SiGe CMOS 0.35\,\textmu m and has surface of 3.9\,mm~x~5.8\,mm = 22\,mm\textsuperscript{2}. It uses a total power of 445\,mW (110\,mA @ 3.3\,V, 55\,mA @ 1.5\,V). 

\begin{figure}[ht]
\begin{center}
  \includegraphics[width=1\textwidth]{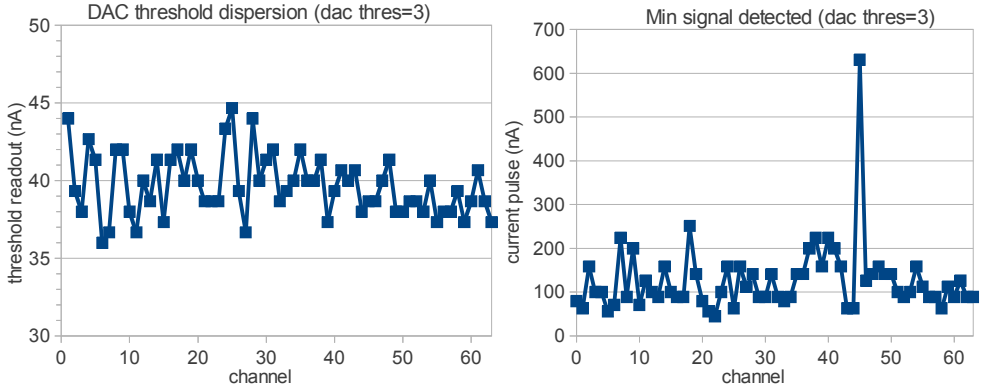}
\caption{Typical DAC threshold and minimal threshold dispersion over one representative ASIC.}
\label{TestDiscriASIC}
\end{center}
\end{figure}

Fig.~\ref{TestDiscriASIC} shows the typical DAC threshold and minimal threshold dispersion over one representative ASIC, where the latter is determined by recording the current pulse amplitude required to trigger a channel at a fixed DAC setting.
The plot shows that the DAC levels are totally homogeneous but that the effective thresholds have a few spikes remaining. This is explained by the offset correction efficiency dispersion between channels, which in turn has an influence on the minimum signal detected at a given threshold.

\section{Readout electronics}
\label{DAQ.sec}
\subsection{Electronic board overview}
\begin{figure}[ht]
\begin{center}
\includegraphics[width=0.8\textwidth]{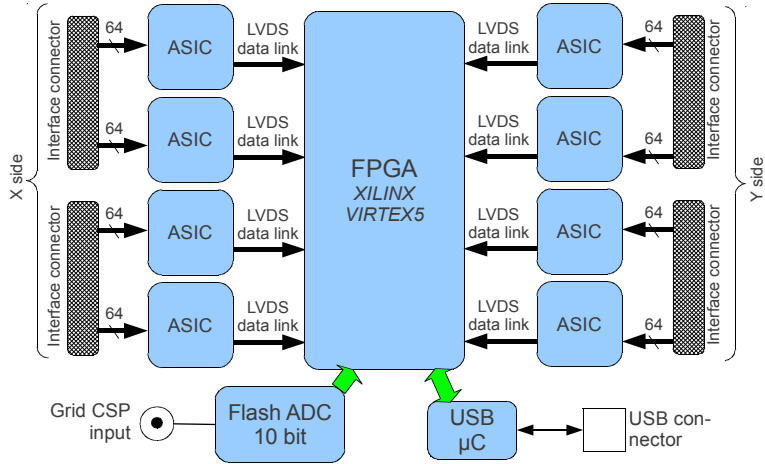}
\caption{Block diagram of the acquisition board. It comprises 8 dedicated ASICs, a FPGA, a flash ADC and an USB interface for DAQ and slow control interface.}
\label{BlockDAQBoard} 
\end{center}
\end{figure}

The readout electronic, which is an upgrade of a previous work (Bourrion~\etal~\cite{Bourrion}), comprises 8 dedicated ASICs, a FPGA, a flash ADC and an USB interface for DAQ and slow control (see fig.~\ref{BlockDAQBoard}).
The connection between the anode located inside the chamber and the electronics at ambient pressure is done via an airtight interface (Igaz~\etal~\cite{Iguaz}). Each strip input is equipped with a discharge protection. The hit strip information, that is sampled at a rate of 50\,MHz, is transferred by 8 LVDS serial links at 400\,MHz to a unique processing FPGA. This FPGA allows the auto-triggering and does the first level event building.

In parallel to the anode signal processing, the grid signal is fed to flash ADC and sampled at 50\,MHz. While keeping a very good energy resolution, an estimation of the charge deposit through time can be obtained off-line by deriving the digitized CSP signal.

The board has a dimension of 25\,cm$\times$25\,cm and uses 9.4\,W in operation. 

\subsection{FPGA firmware}
\begin{figure}[ht]
\begin{center}
 \includegraphics[width=0.95\textwidth]{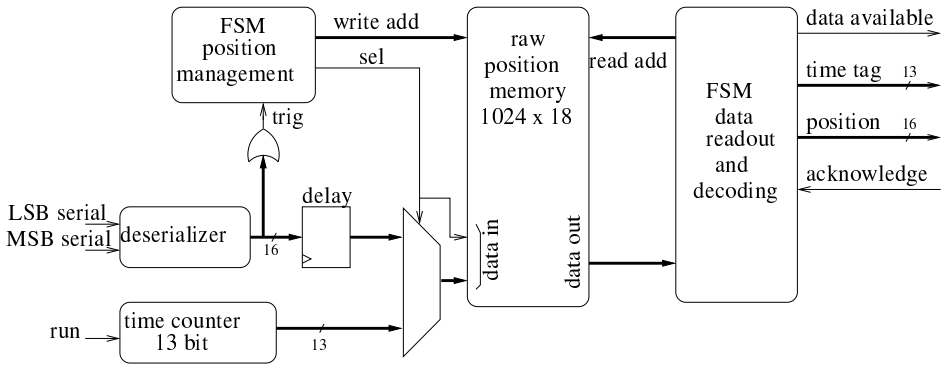}
\caption{Block diagram of the ASIC interface. It comprises the data deserializer, the local trigger building, intermediate buffering and the first level processing.}
\label{GroupManagement}
\end{center}
\end{figure}
As shown in fig.~\ref{GroupManagement}, the FPGA deserializes the data received from the ASIC and for each group of 16 channels a local trigger is built (OR). The first level processing starts at this stage, i.e. when a coincidence exists between X and Y strips, a local recording  takes place (start date + positions). In theory, a single event would be defined as continuously firing strips. Unfortunately, the primary electron distribution can be noncontinuous, therefore untriggered strips can split the track (clusters). To cope with this, the recording is actually stopped when there are no more fired strips for a preset number of clock cycles.

Most of the time a few strips only are fired, hence by using an adequate data encoding (shown in table~\ref{TableEnc}), the data payload per event can be reduced. For instance, when 2  strips in X and 2 strips in Y are fired in the same time slice, taking advantage of the encoding, only 64 bit are transfered instead of 512. 
\begin{table}
\begin{center}
\begin{small}
\begin{tabular}{|l|c|c|c|c|c|c|}
\hline
  \emph{Bit \#} & [15..13] & 12 & [11..10] & [9..8] & [7..4] & [3..0]\\
\hline
\emph{Content} & 0 & X or Y & ASIC\# & Group\# & 0 & ch\# \\
\hline
\end{tabular}
\end{small}
\end{center}
\caption{Position data encoding.}
\label{TableEnc}
\end{table}
This first encoding and processing stage is done in parallel for the X and Y side. At the following stage, dedicated state machines search and aggregate data from the same time slice in order to perform the first level event building. This association is done in several stages, in order to concentrate more and more the data, and to present a single buffer to the USB interface.

The energy measurement is done in parallel to the position processing. The CSP signal recording is armed by the position triggering, but given the fact that the signal path delay is different for the anode and the grid signal, the actual recording is performed when the grid integrated signal is digitized. 
As a consequence of the low level signal compared to the noise level, a slope condition, which is more robust and noise immune than using a simple level threshold, is used. Taking advantage of the FPGA, dedicated filtering techniques are implemented (CIC and FIR filters) to further enhance the signal to noise ratio without degrading the event signal shape.

Consequently to the parallel processing, the position and energy data are recorded in separate FIFOs for USB readout.

\section{Acquisition software}
\label{soft.sec}
The first task of the acquisition software is to re-associate the position and the energy data. It uses the position information which is provided in a list of X/Y coordinates fired per time slice. Knowing that an event is defined as continuously triggering strips, the algorithm basically searches continuous triggering position (in time) and searches a discontinuity (time tag jumps by more than the preset value) to close the event. Once the event is defined, the energy information with a corresponding time tag is associated.
The event building is then finished and contains directly the coordinates of the fired strips coordinate for each time slice and the corresponding digitized grid signal.

\begin{figure}[ht]
\begin{center}
\includegraphics[width=0.95\textwidth]{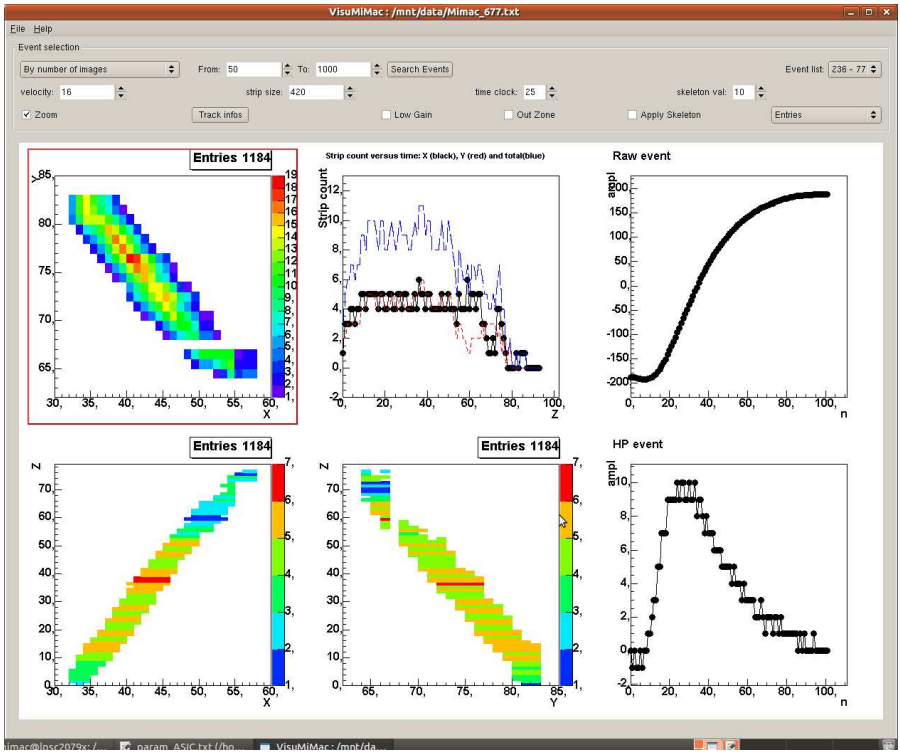} 
\caption{Screenshot of the real time display. The 3 projections of the track (XY-XZ-ZY), the number of fired strips per time slice, the grid digitized signal and its derivative can be seen.}
\label{EventDisplay}
\end{center}
\end{figure}

The second task of the acquisition software is to provide a real time display (see fig.~\ref{EventDisplay}). The Graphical User Interface (GUI) provides 3 projections of the track (XY-XZ-YZ), the number of fired strips per time slice, the Track CSP digitized signal and its derivative. It also offers the possibility to search  (energy, duration, ...) and display a specific recoil track. An energy histogram of the run is also built on line and displayed on an other tab (not shown in fig.~\ref{EventDisplay}).

\section{Summary}
\label{SummarySec}
A complete dedicated solution from front-end to back-end was developed to instrument a MIMAC prototype ($256 \times 256$). Several MIMAC electronics can be connected per computer and the event building developed can be performed in several stages provided a common time tagging electronic board is designed and a synchronization upgrade is implemented.


\end{document}